\documentclass[aps,prb,reprint,superscriptaddress,showpacs,floatfix]{revtex4-1}
\usepackage{amsmath}
\usepackage{graphicx,longtable}
\usepackage{bm}
\usepackage{dcolumn}

\begin{document}

\title{\boldmath Spin dynamics and spin freezing\\ in the triangular lattice antiferromagnets FeGa$_2$S$_4$ and NiGa$_2$S$_4$}

\author{Songrui Zhao}
\altaffiliation[Present address: ]{Department of Electrical and Computer Engineering, McGill University, Montreal, Quebec, Canada H3A 2A7.}
\affiliation{Department of Physics and Astronomy, University of California, Riverside, California 92521, USA}
\author{P. Dalmas de R\'eotier}
\affiliation{Institut Nanosciences et Cryog\'enie, SPSMS, CEA\\ and Universit\'e Joseph Fourier, F-38054 Grenoble, France}
\author{A. Yaouanc}
\altaffiliation[Present address: ]{Laboratory for Muon-Spin Spectroscopy, Paul Scherrer Institute, 5232 Villigen-PSI, Switzerland}
\affiliation{Institut Nanosciences et Cryog\'enie, SPSMS, CEA\\ and Universit\'e Joseph Fourier, F-38054 Grenoble, France}
\author{D. E. MacLaughlin}
\affiliation{Department of Physics and Astronomy, University of California, Riverside, California 92521, USA}
\author{J. M. Mackie}
\affiliation{Department of Physics and Astronomy, University of California, Riverside, California 92521, USA}
\author{O. O. Bernal}
\affiliation{Department of Physics and Astronomy, California State University, Los Angeles, California 90032, USA}
\author{Y. Nambu}
\altaffiliation[Present address: ]{Neutron Science Laboratory, ISSP, 106-1 Shirakata, Tokai 319-1106, Japan.}
\affiliation{Institute for Solid State Physics, University of Tokyo, Kashiwa 277-8581, Japan}
\author{T. Higo}
\affiliation{Institute for Solid State Physics, University of Tokyo, Kashiwa 277-8581, Japan}
\author{S. Nakatsuji}
\affiliation{Institute for Solid State Physics, University of Tokyo, Kashiwa 277-8581, Japan}

\date{\today}

\begin{abstract}
Magnetic susceptibility and muon spin relaxation ($\mu$SR) experiments have been carried out on the quasi-2D triangular-lattice spin $S = 2$ antiferromagnet~FeGa$_2$S$_4$. The $\mu$SR data indicate a sharp onset of a frozen or nearly-frozen spin state at $T^* = 31(2)$~K, twice the spin-glass-like freezing temperature~$T_f = 16(1)$~K\@. The susceptibility becomes field dependent below $T^*$, but no sharp anomaly is observed in any bulk property. A similar transition is observed in $\mu$SR data from the spin-1 isomorph~NiGa$_2$S$_4$. In both compounds the dynamic muon spin relaxation rate~$\lambda_d(T)$ above $T^*$ agrees well with a calculation of spin-lattice relaxation by Chubukov, Sachdev, and Senthil in the renormalized classical regime of a 2D frustrated quantum antiferromagnet. There is no firm evidence for other mechanisms. At low temperatures $\lambda_d(T)$ becomes temperature independent in both compounds, indicating persistence of spin dynamics. Scaling of $\lambda_d(T)$ between the two compounds is observed from ${\sim}T_f$ to ${\sim}1.5T^*$. Although the $\mu$SR data by themselves cannot exclude a truly static spin component below $T^*$, together with the susceptibility data they are consistent with a slowly-fluctuating ``spin gel'' regime between $T_f$ and $T^*$. Such a regime and the absence of a divergence in $\lambda_d(T)$ at $T^*$ are features of two unconventional mechanisms: (1)~binding/unbinding of $Z_2$ vortex excitations, and (2)~impurity spins in a nonmagnetic spin-nematic ground state. The absence of a sharp anomaly or history dependence at $T^*$ in the susceptibility of FeGa$_2$S$_4$, and the weakness of such phenomena in NiGa$_2$S$_4$, strongly suggest transitions to low-temperature phases with unconventional dynamics.

\end{abstract}


\maketitle

\section{Introduction} \label{sec:intro} 

Among geometrically frustrated magnets the simplest example is the two-dimensional (2D) triangular-lattice Heisenberg antiferromagnet (THAFM) with isotropic spin-spin interactions. This is the system for which a spin liquid state in more than one dimension was first proposed by Anderson\cite{Ande73} as a quantum disordered state in which long-range magnetic order is destroyed. It is now believed that the spin{-}1/2 2D THAFM with nearest-neighbor coupling orders at $T = 0$ with a 120$^{\circ}$ spin structure.\cite{HuEl88,BLP92,CTS99} A number of treatments of the 2D THAFM at nonzero temperatures have been reported.\cite{[{For reviews see }] [{, particularly A.~M. L\"auchli, \textit{ibid.} Chap.~18, and references therein.}] LMM11} Phase transitions associated with the value of the spin~$S$, topological defects, interactions beyond nearest neighbors, additional terms in the Hamiltonian, etc., have been studied intensively. One such transition involves binding and unbinding of topological excitations, the so-called $Z_2$ vortices.\cite{KaMi84,KaKi93,KaYa07,KYO10,Kawa11} Integer spins and additional biquadratic nearest-neighbor coupling are predicted to yield spin-nematic phases with no dipolar ordering.\cite{TsAr06,[{For a review see }] PeLa11inbib} Coupling beyond nearest neighbors may lead to a quantum spin disordered ground state\cite{MWI02} or a state with broken $C_3$ symmetry.\cite{TaKa08,STB09} It has proved difficult to find experimental evidence for many of these transitions, although $Z_2$ vortex binding has been proposed for a number of candidate 2D THAFM systems\cite{Kawa11} and a recent theory\cite{TaTs11} invokes impurity-spin interactions in a nonmagnetic spin-nematic low-temperature phase.

The quasi-2D triangular-lattice antiferromagnetic insulator~NiGa$_2$S$_4$ was characterized by Naka\-tsuji and co-workers\cite{NNTS05,[{For a review and references through 2009 see }] NNO10} and found to possess a number of unusual properties that have attracted considerable attention, both experimentally\cite{NTON07,TIKI08,YKHN08,YKHN10,YDdRCM08,NNMO08,MNNH08,MNNI09,DdRYCM09,%
MNOM10,Namb08,MKK08,SJBN10,TNNW10,MKLK10,MKKK11} and the\-o\-ret\-i\-cal\-ly.\cite{LMP06,BSS06,TsAr07,KaYa07,Mazi07,Cher08,TaKa08,PoKi09,STB09,%
KYO10,LZN10,Kawa11,TaTs11} It is a 2D THAFM with Ni$^{2+}$ spin~$S = 1$ and very little spin anisotropy. Third-nearest-neighbor antiferromagnetic exchange is dominant, with a weak nearest-neighbor ferromagnetic interaction\cite{NNTS05,YKHN08} as suggested by calculated superexchange pathways.\cite{TMSN07} A phase transition, possibly influenced by impurity effects, is indicated by a cusp and weak bifurcation between field-cooled (FC) and zero-field-cooled (ZFC) dc magnetic susceptibilities~$\chi_\mathrm{dc}(T)$ at $T^* = 8.5$--9~K\@.\cite{NNO10} 

A drastic slowing of Ni$^{2+}$ spin fluctuations as $T^*$ is approached from above is observed in NMR,\cite{TIKI08} ESR,\cite{YKHN08,YKHN10} and muon spin relaxation ($\mu$SR)\cite{YDdRCM08,MNNH08,MNNI09,DdRYCM09,MNOM10} experiments, with a frozen or nearly-frozen Ni$^{2+}$ spin state below $T^*$ characterized by an order-parameter-like temperature dependence of the (nearly) frozen moment and strong spin fluctuations down to 25~mK\@. The frequency-dependent ac magnetic susceptibility~$\chi_\mathrm{ac}(T)$ suggests a spin-glass-like transition at a lower temperature~$T_f = 2.2$--2.7~K.\cite{Namb08} Based on these results, a viscous ``spin-gel'' or extended thermodynamic critical regime has been proposed between ${\sim}T_f$ and $T^*$.\cite{Namb08,NNO10} A similar broad fluctuating regime is observed in NaCrO$_2$.\cite{OMBU06} 

Perhaps the most mysterious property of NiGa$_2$S$_4$ is the magnetic specific heat~$C_M$, which is independent of applied magnetic field up to 7~T and exhibits a $T^2$ temperature dependence at low temperatures.\cite{NNTS05} This implies linearly-dispersing 2D excitations that do not couple to the field, which seems hard to reconcile with the observation that the low-temperature muon spin relaxation rate is rapidly suppressed by applied field.\cite{MNNH08} 
 
The isostructural insulator~FeGa$_2$S$_4$,\cite{NTON07,NNO10} in which Fe$^{2+}$ is in the $t_{2g}^{4}e_{g}^{2}$ high-spin $S = 2$ configuration, is also a candidate for a 2D THAFM\@. Both FeGa$_2$S$_4$ and NiGa$_2$S$_4$ exhibit FC-ZFC bifurcation at ${\sim}|\theta_W|/10$, where $\theta_W$ is the (negative) Weiss temperature from the paramagnetic-state susceptibility at high temperatures. Normally bifurcation occurs at a spin freezing temperature~$T_f$. Thus the Ramirez parameter\cite{Rami94}~$\theta_W/T_f$ is large (${\gtrsim}\,10$), indicating strong frustration. In both compounds $C_M$ exhibits an unusual two-peak temperature dependence with no sign of a phase transition. As in NiGa$_2$S$_4$, $C_M(T)$ in FeGa$_2$S$_4$ follows a field-independent $T^2$ law at low temperatures. Some properties of the two materials differ significantly, however. In NiGa$_2$S$_4$ $\chi_\mathrm{ac}(T)$ exhibits a frequency dependence below $T_f$,\cite{NNO10} whereas in FeGa$_2$S$_4$ bifurcation, much stronger than that at $T^*$ in NiGa$_2$S$_4$, is observed at $T_f \approx 16$~K,\cite{NTON07} indicating a spin-glass-like transition at this temperature. In the Ni$_{1-x}$Fe$_x$Ga$_2$S$_4$ series the specific heats scale with $\theta_W$ in the $T^2$ region but not at higher temperatures.\cite{NNMO08}

To date FeGa$_2$S$_4$ has been studied using only a few microscopic techniques. In Ni$_{1-x}$Fe$_x$Ga$_2$S$_4$, $0.01 \le x \le 1$, $^{57}$Fe M\"ossbauer-effect (ME) measure\-ments\cite{MKK08,MKLK10} indicate spin freezing at a temperature~$T^*$ that varies smoothly between 12(1)~K for $x = 0.01$ and $33(1)~\mathrm{K} \approx 2T_f$ for $x=1$,\cite{MKLK10} with an order-parameter-like temperature dependence of the frozen moment and strong spin fluctuations below $T^*$ similar to the $\mu$SR results in NiGa$_2$S$_4$. Initial $\mu$SR experiments in FeGa$_2$S$_4$\cite{DdRYMZ12} revealed a magnetic transition at and strong muon relaxation below $T^* \approx 30$~K, confirming the ME results. It was noted that the relation of $T^*$ to $C_M(T)$ is not the same in the two compounds: in NiGa$_2$S$_4$ $T^*$ lies just below the low-temperature peak in $C_M(T)$, whereas in FeGa$_2$S$_4$ $T^*$ is found between the two peaks. As a result $C_M(T)$ does not scale with $T^*$, and the relation between spin freezing and entropy release is unclear. 

This paper (1)~reports further measurements of $\chi_\mathrm{dc}(T)$ and results of $\mu$SR experiments in FeGa$_2$S$_4$, which were undertaken to clarify the nature of spin freezing and to probe Fe$^{2+}$ spin dynamics, and (2)~compares $\mu$SR and other data from FeGa$_2$S$_4$ with those from NiGa$_2$S$_4$. The susceptibility data confirm earlier results,\cite{NTON07} and extend measurements to applied fields in the range~0.1--2~T\@. As previously noted,\cite{DdRYMZ12} there is no anomaly in the muon relaxation rate at $T_f$ and also no sharp anomaly in $\chi_\mathrm{dc}$ at $T^*$; the latter is an extremely unusual feature of this compound,\cite{[{In spin glasses, for example, $\chi(T)$ exhibits a strong well-defined cusp and bifurcation at $T_f$ even though $C_M(T)$ is smooth there; see, e.g., }] Mydo93} not shared by NiGa$_2$S$_4$. 

The $\mu$SR results from FeGa$_2$S$_4$ provide new insight into spin freezing and dynamical fluctuations in these compounds. The data reveal an abrupt transition at $T^* \approx 31$~K to a phase with frozen or nearly-frozen spins and strong spin fluctuations but with almost no signature in bulk measurements. The agreement with the ME results rules out perturbation by the muon electric charge\cite{Varm08} as the origin of this behavior.

The theory of spin-wave excitations in a 2D frustrated quantum antiferromagnet has been treated by Chubukov, Sachdev, and Senthil\cite{CSS94a,CSS94b,CSS94c} (hereafter CSS) and others.\cite{ADM92,LBLP95} Comparison of muon relaxation rates above $T^*$ in both compounds with the CSS result for the spin-lattice relaxation rate of a probe spin (nucleus or muon) yields values of the $3d$ exchange interaction in remarkably good agreement with previous studies.\cite{NNTS05,NTON07,YKHN08,SJBN10} There is no firm evidence for muon relaxation due to other mechanisms.\cite{STB09,KYO10,Kawa11} 

The muon relaxation rates in FeGa$_2$S$_4$ scale with those in NiGa$_2$S$_4$\cite{MNNH08} over a wide temperature range from ${\sim}T_f$ to ${\sim}1.5T^*$. Although the $\mu$SR data by themselves do not rule out a truly static spin component, as found in conventional magnetic phases, the spin dynamics revealed by $\mu$SR together with the absence of a susceptibility anomaly at $T^*$ are consistent with an extended critical regime or slowly-fluctuating ``spin-gel'' state in both compounds below $T^*$. 

\section{Experimental Procedure and Results} \label{sec:expt}

\subsection{Magnetic susceptibility}

A powder sample of FeGa$_2$S$_4$ was prepared as described previously.\cite{D-SDP80,NNTS05,NMHI09} The bulk dc susceptibility was measured using a Quantum Design Magnetic Properties Measurements System over the temperature range~2--300~K for applied magnetic fields in the range~0.1--6~T\@. For FC measurements the magnetic field was set and the sample was cooled to 2~K before collecting the data, while for ZFC measurements the sample was cooled in zero field, the field was set at 2~K, and data were taken upon warming. After each measurement the sample was warmed to 350~K in zero field to quench any magnetic order. 

\begin{figure}[ht]
\includegraphics[clip=,width=8cm]{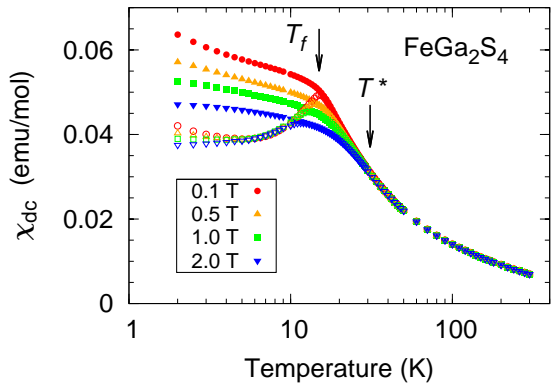}
\caption{\label{fig:suscept} (Color online) Temperature dependence of the dc magnetic susceptibility~$\chi_\mathrm{dc}$ of FeGa$_2$S$_4$ for various applied magnetic fields. Filled symbols: field-cooled (FC) data. Open symbols: zero-field-cooled (ZFC) data. $T_f$: low-field spin freezing temperature. $T^*$: onset temperature of quasistatic muon spin relaxation [cf.\ Fig.~\protect\ref{fig:rlx20G}(a)].}
\end{figure}

The results are shown in Fig.~\ref{fig:suscept}. 
Consistent with a previous report,\cite{NTON07} for applied field~$\mu_0H = 0.1$~T bifurcation between the FC and ZFC data is observed below 16(1)~K, suggesting spin-glass-like freezing at this temperature. The data indicate a strong suppression of the spin freezing temperature with applied field, similar to that observed in Fe-rich (Ni,Fe)Ga$_2$S$_4$ alloys,\cite{NaNa11} to about 10~K at 2~T; this is also consistent with spin-glass-like freezing. As in NiGa$_2$S$_4$,\cite{Namb08} the susceptibility is suppressed by field (i.e., a negative nonlinear susceptibility sets in) for temperatures $\lesssim T^*$ obtained from ME and $\mu$SR experiments [cf.\ Fig.~\ref{fig:rlx20G}(a) below], but there is no sharp feature in $\chi_\mathrm{dc}(T)$ at $T^*$. As noted in Sec.~\ref{sec:intro}, the bifurcation in FeGa$_2$S$_4$ is much stronger than that in NiGa$_2$S$_4$. 

\subsection{Muon spin relaxation} \label{subsec:musr}

The $\mu$SR technique is a sensitive probe of static and dynamic magnetism in magnetic materials.\cite{Sche85,YaDdR11} In $\mu$SR experiments spin-polarized positive muons are implanted in the sample and come to rest at interstitial sites. Each muon precesses in the sum of the local field~$\mathbf{B}_\mathrm{loc}(t)$ due to its magnetic environment and any applied field, and decays with a mean lifetime~$\tau_\mu = 2.197~\mu$s into a positron and two neutrinos. The positron is emitted preferentially in the direction of the muon spin at the time of decay, so that detection of decay positrons permits determination of the evolution of the muon spin polarization. The resulting signal can be observed up to times of several $\tau_\mu$ (typically 10--15~$\mu$s) and is conceptually similar to the free induction decay of NMR,\cite{Slic96} although the detection technique is very different. 
 
The onset of a quasistatic\footnote{A component~$\langle\mathbf{B}_\mathrm{loc}\rangle$ of the muon local field is quasistatic if it fluctuates at a slow rate compared to the muon precession frequency in $\langle B_\mathrm{loc}\rangle$. We include the static limit in our use of this term.} component~$\langle\mathbf{B}_\mathrm{loc}\rangle$ of $\mathbf{B}_\mathrm{loc}(t)$ is expected at a magnetic phase transition. In general muon spin relaxation [decay of the ensemble muon spin polarization~$P(t)$] is due to a combination of two relaxation mechanisms:\cite{Sche85,YaDdR11}
\begin{itemize}
 
\item \textit{quasistatic} relaxation, due to dephasing of muon spin precession in $\langle\mathbf{B}_\mathrm{loc}\rangle$ if its magnitude is inhomogeneously distributed, and

\item \textit{dynamic} (spin-lattice) relaxation of the muon spin due to thermal fluctuations of $\mathbf{B}_\mathrm{loc}(t)$.

\end{itemize}
These are of course the inhomogeneous and homogeneous relaxation mechanisms of NMR.
 
It is straightforward to separate the relaxation rates associated with these processes in $\mu$SR experiments in zero and weak longitudinal magnetic field~$H_L$,\cite{Sche85,YaDdR11} provided that the dynamic muon relaxation is slower than the quasistatic muon relaxation.\cite{KuTo67,*HUIN79} Then $P(t)$ has a characteristic two-component form that can be modeled by
\begin{equation}
P(t) = (1-\eta_d)P_q(t) + \eta_dP_d(t) \,.
\label{eq:2cmpnt}
\end{equation}
Here the first and second terms on the right-hand side describe quasistatic and dynamic relaxation processes that control the evolution of $P(t)$ at early and late times, respectively,\cite{KuTo67,HUIN79,YaDdR11} and $\eta_d$ is the fraction of the initial muon spin polarization that is relaxed dynamically, i.e., that survives at late times after $P_q(t) \to 0$. Observation of such two-component behavior indicates the presence of quasistatic magnetism, with or without long-range order. In a randomly-oriented powder sample $\eta_d = 1/3$ for $\langle B_\mathrm{loc}\rangle \gg \mu_0H_L$ and $\eta_d \to 1$ for $\langle B_\mathrm{loc}\rangle \ll \mu_0H_L$,\cite{HUIN79} so that for zero or low $H_L$ a sudden decrease of $\eta_d$ with decreasing temperature signals the onset of a quasistatic local field due to a magnetic phase transition.

$\mu$SR experiments were carried out on a powder sample of FeGa$_2$S$_4$ using the M20 beam line at TRIUMF, Vancouver, Canada. The asymmetry~$A(t)$ in positron count rate, which is proportional to $P(t)$, was measured using the standard time-differential $\mu$SR technique.\cite{Sche85,YaDdR11} Data were taken from 2~K to 260~K in a weak longitudinal field~$\mu_0H_L \approx 2$~mT to decouple nuclear dipolar fields,\cite{HUIN79} and also for $\mu_0H_L$ between 2 and 100~mT at 1.7~K\@. 

\begin{figure}[t]
\includegraphics[clip=,width=8cm]{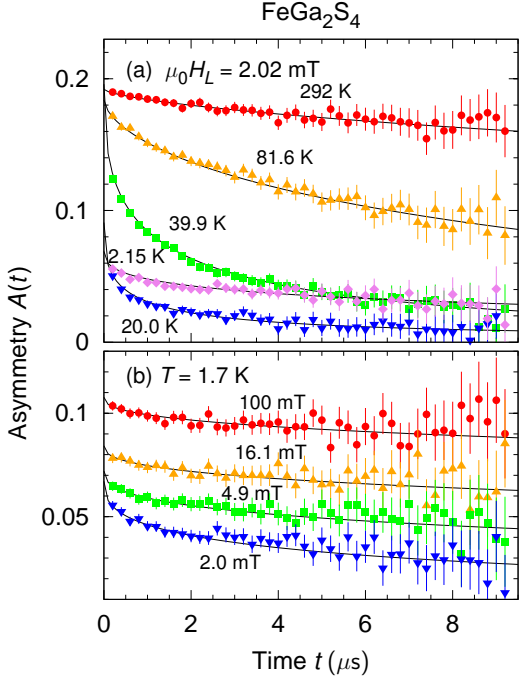}
\caption{\label{fig:asy} (Color online) Representative late-time muon asymmetry data (positron count asymmetry vs time) in FeGa$_2$S$_4$. (a)~Asymmetry vs time at various temperatures for longitudinal field~$\mu_0H_L = 2.02$~mT\@. (b)~Asymmetry vs time for various values of $\mu_0H_L$ at temperature~$T = 1.7$~K\@. Curves: fits to Eq.~(\protect\ref{eq:rlxfn}).}
\end{figure}

Representative late-time muon asymmetry data for $\mu_0H_L = 2.02$~mT are shown in Fig.~\ref{fig:asy}(a). 
As previously reported,\cite{DdRYMZ12} the initial asymmetry decreases rapidly between 40 and 20~K\@. This indicates\cite{HUIN79} the onset of quasistatic Fe$^{2+}$ spin freezing (reduction of $\eta_d$) as discussed above. As noted in Sec.~\ref{sec:intro}, ME studies\cite{MKK08,MKLK10} of Ni$_{1-x}$Fe$_x$Ga$_2$S$_4$ give evidence for quasistatic Fe$^{2+}$ spins below $T^*$, consistent with this loss of asymmetry. Each technique places a lower limit on the correlation time~$\tau_c$ of $\langle\mathbf{B}_\mathrm{loc}\rangle$. In ME and $\mu$SR experiments these limits are ${\sim}10^{-8}$~s and ${\sim}10^{-7}$--$10^{-6}$~s (cf.\ Sec.~\ref{subsec:below}), respectively, so that the $\mu$SR results are a somewhat stronger. 

Unfortunately the early-time quasistatic muon relaxation could not be observed at low temperatures in FeGa$_2$S$_4$,\cite{DdRYMZ12} because the initial decay time was shorter than the spectrometer ``dead time'' between a muon stop and the earliest detection of the decay positron ($\sim$10~ns). Thus $\langle B_\mathrm{loc}\rangle$ is large and broadly distributed. The spectrometer dead time puts a lower limit of roughly $200~\mu\mathrm{s}^{-1}$ on the quasistatic muon relaxation rate~$\Lambda_q$, corresponding to a quasistatic local field distribution width~$\langle B_\mathrm{loc}\rangle_\mathrm{rms} = \Lambda_q/\gamma_\mu \gtrsim 0.2$~T; here $\gamma_\mu = 851.56~\mu\mathrm{s}^{-1}~\mathrm{T}^{-1}$ is the muon gyromagnetic ratio. The average local field~$\langle B_\mathrm{loc}\rangle_\mathrm{av}$ may be considerably larger than this; in NiGa$_2$S$_4$, where the early-time asymmetry was observable, $\langle B_\mathrm{loc}\rangle_\mathrm{av} \gtrsim 5\,\langle B_\mathrm{loc}\rangle_\mathrm{rms}$ (Refs.~\onlinecite{YDdRCM08,MNNH08}).

The asymmetry data were fit to a relaxation function of the form
\begin{equation}
A(t) = A_0P(t)\,,
\label{eq:rlxfn}
\end{equation} 
where $A_0$ is the initial count-rate asymmetry and $P(t)$ is given by Eq.~(\ref{eq:2cmpnt}) with
\begin{equation}
 P_d(t) = \exp[-(\lambda_d t)^\beta] \,,
\label{eq:dynrlxfn}
\end{equation}
the stretched-exponential form with relaxation rate~$\lambda_d$ and stretching power~$\beta < 1$. This is a convenient parametrization of sub-exponential relaxation due to an inhomogeneous distribution of locally exponential relaxation processes.\cite{John06,YaDdR11} $\lambda_d$ is a characteristic relaxation rate (not the average),\footnote{From Eq.~(\ref{eq:dynrlxfn}) $P_d(t) = 1/e$ for $\lambda_d t = 1$ independently of $\beta$.} and $\beta$ controls the width of the distribution, which becomes larger for smaller $\beta$. Representative fits are shown in Fig.~\ref{fig:asy}. The stretched-exponential form was also used to analyze late-time data from $\mu$SR experiments in NiGa$_2$S$_4$ using powder samples\cite{YDdRCM08,MNNH08,MNNI09,MNOM10} and a mosaic of oriented single crystals with $\mathbf{H}_L$ normal to the Ni planes.\cite{MNNH08} The origin of this relaxation-rate inhomogeneity is not clear, since structural studies of these materials found nearly perfect triangular NiS$_2$ lattice layers with no distortion.\cite{NNTS05,NMHI09}  It does not seem to be the stacking faults inferred from NQR,\cite{TIKI08,NMHI09} since the bimodal distribution this would produce is not observed. We note that spin disorder and a short spin-spin correlation length are found in neutron scattering experiments on NiGa$_2$S$_4$.\cite{NNTS05,MNNH08}

\begin{figure}[t]
\includegraphics[clip=,width=8cm]{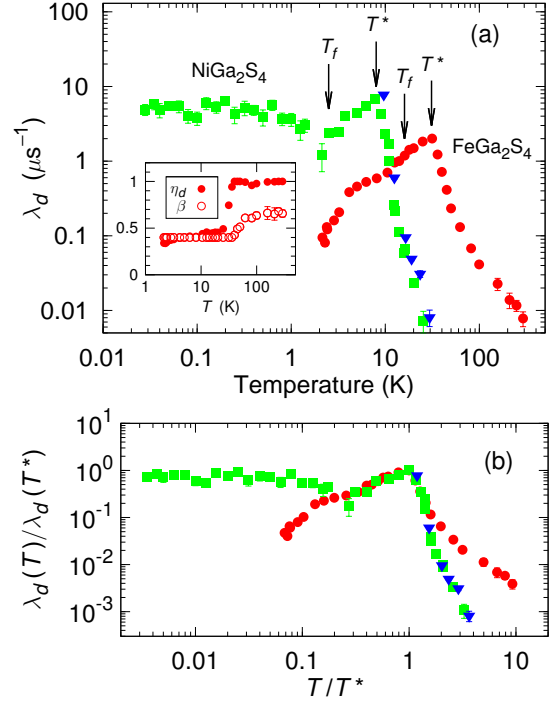}
\caption{\label{fig:rlx20G} (Color online) (a)~Temperature dependence of dynamic muon spin relaxation rate $\lambda_d$ in weak longitudinal applied field~$H_L$ in FeGa$_2$S$_4$ and NiGa$_2$S$_4$. Circles: FeGa$_2$S$_4$ powder, $\mu_0H_L = 2.02$~mT\@. Squares: NiGa$_2$S$_4$ powder, $\mu_0H_L = 2.0$~mT\@. Triangles: NiGa$_2$S$_4$, mosaic of oriented single crystals, $\mathbf{H}_L \parallel \mathbf{c}$, $\mu_0H_L = 2.0$~mT. $T_f(\mathrm{FeGa_2S_4})$: freezing temperature from FC-ZFC bifurcation in $\chi_\mathrm{dc}(T)$ (Fig.~\protect\ref{fig:suscept}). $T_f(\mathrm{NiGa_2S_4)}$: freezing temperature from onset of frequency-dependent $\chi_\mathrm{ac}(T)$ (Refs.~\protect\onlinecite{Namb08,NNO10}). $T^*$: transition temper\-a\-tures from $\mu$SR data. Inset: FeGa$_2$S$_4$, late-time asymmetry fraction~$\eta_d$ [Eq.~(\protect\ref{eq:2cmpnt})] (filled circles) and stretching power~$\beta$ [Eq.~(\protect\ref{eq:dynrlxfn})] (open circles). (b)~Normalized muon relaxation rate~$\lambda_d(T)/\lambda_d(T^*)$ vs normalized temperature~$T/T^*$. Symbols as in (a).}
\end{figure}

The temperature dependencies of $\lambda_d$ for $\mu_0H_L \approx 2$~mT in FeGa$_2$S$_4$ and NiGa$_2$S$_4$ are shown in Fig.~\ref{fig:rlx20G}(a).\footnote{The muon lifetime normally limits measurable muon relaxation rates to $\gtrsim 10^{-2}\ \mu\mathrm{s}^{-1}$.} 
Sharp cusps in $\lambda_d(T)$ are observed at $T^* = 31(2)$~K and 8.5(1.0)~K in FeGa$_2$S$_4$ and NiGa$_2$S$_4$, respectively. In both compounds $\lambda_d(T)$ increases by more than two orders of magnitude over a wide temperature range as $T \to T^*$ from above. This is clear evidence for critical slowing down of magnetic fluctuations: as in NMR, $\lambda_d$ is proportional to the spin correlation time in the motional narrowing limit appropriate to the paramagnetic state.\cite{Slic96,YaDdR11} From Fig.~\ref{fig:rlx20G}(a), in FeGa$_2$S$_4$ $T^*$ is about twice the spin freezing temperature $T_f$.

For $T < T^*$ $\lambda_d(T)$ decreases with decreasing temperature in both compounds but saturates and shows no sign of vanishing as $T \to 0$ (down to $T/T^* \approx 0.003$ in FeGa$_2$S$_4$\cite{DdRYMZ12}). Such persistence of strong spin dynamics to low temperatures is a general feature of geometrically frustrated magnets\cite{DKCG96,*GGG10,*MCDMG11,*CaKe11inbib} and is not well understood, but seems to indicate a zero-energy singularity in the density of excited states.\cite{YDdRGM05} In FeGa$_2$S$_4$ there is no anomaly in $\lambda_d(T)$ at $T_f$. This is very surprising, and is not understood: the observed strong FC-ZFC bifurcation is usually considered good evidence for spin freezing, which in turn would be expected to reduce the relaxation rate. In NiGa$_2$S$_4$ a minimum in $\lambda_d(T)$ is seen near $T_f$, but data above and below this temperature were taken in different cryostats and an experimental artifact cannot be ruled out. 
 
The temperature dependencies of the late-time asymmetry fraction~$\eta_d$ and the stretching power~$\beta$ in FeGa$_2$S$_4$ are shown in the inset to Fig.~\ref{fig:rlx20G}(a). At low temperatures $\eta_d \approx 0.3$--0.5, close to the value~1/3 expected for $\mu_0H_L \ll \langle B_\mathrm{loc}\rangle$, and then rises rapidly to $\sim$1 at $T^*$. This is strong evidence for frozen or nearly-frozen Fe$^{2+}$ spins below $T^*$, since it is the expected behavior if $\langle \mathbf{B}_\mathrm{loc}\rangle$ sets in suddenly at this temperature.

Below $T^*$ $\beta$ is considerably smaller than 1, although it is not accurately determined by the data; it was necessary to constrain $\beta$ in this region to obtain consistent fits.\cite{DdRYMZ12} We set a minimum~$\beta = 0.4$, but the behavior of $\lambda_d(T)$ is not very sensitive to the specific choice. For $T > T^*$ $\beta$ tends to be larger but still $< 1$. The stretching has been found to be less pronounced above $T^*$ at other applied fields in both FeGa$_2$S$_4$ and NiGa$_2$S$_4$.\cite{YDdRCM08,DdRYCM09,DdRYMZ12}

Figure~\ref{fig:rlx20G}(b) shows the dependence of the normalized muon relaxation rate~$\lambda_d(T)/\lambda_d(T^*)$ on the normalized temperature~$T/T^*$ in FeGa$_2$S$_4$ and NiGa$_2$S$_4$. It can be seen that the data scale from ${\sim}T_f$ to ${\sim}1.5T^*$. This behavior and the possibility of slow quasistatic fluctuations below $T^*$ (cf.\ Sec.~\ref{subsec:below}) suggest an extended critical region, followed by disordered freezing at $T_f$.\cite{Namb08,NNO10} Scaling does not hold for the transition temperatures themselves, however: the ratio~$T^*/|\theta_W|$ is 0.19(1) for FeGa$_2$S$_4$ and 0.11(1) for NiGa$_2$S$_4$.

\begin{figure}[t]
\includegraphics[clip=,width=8cm]{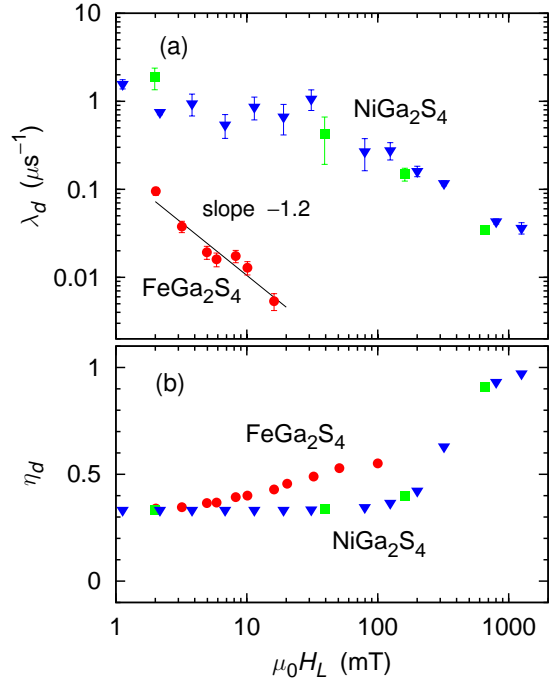}
\caption{\label{fig:rlx2K} (Color online) Dependence of dynamic muon spin relaxation parameters on longitudinal field~$H_L$ in FeGa$_2$S$_4$ and NiGa$_2$S$_4$ at low temperatures ($T \ll T^*$). Circles: FeGa$_2$S$_4$ powder, $T = 1.7$~K\@. Squares: NiGa$_2$S$_4$ powder, $T = 2.1$~K\@. Triangles: NiGa$_2$S$_4$ oriented single-crystal mosaic, $T = 2.3$~K\@. (a)~Relaxation rate~$\lambda_d$. (b)~Late-time asymmetry fraction~$\eta_d$.}
\end{figure}

The field dependencies of $\lambda_d$ and $\eta_d$ in NiGa$_2$S$_4$ and FeGa$_2$S$_4$ for $T \sim 2$~K are shown in Fig.~\ref{fig:rlx2K}.
It can be seen that in both compounds the muon relaxation is suppressed by field. For FeGa$_2$S$_4$ $\lambda_d$ varies roughly as ${H_L}^{-1.2}$, and is suppressed by an order of magnitude for $\mu_0H_L \approx 10$~mT\@. For fields an order of magnitude higher than this $\eta_d$ remains in the neighborhood of 0.4--0.5, close to the value~1/3 expected for $\mu_0H_L \ll \langle B_\mathrm{loc}\rangle$.\cite{HUIN79} In NiGa$_2$S$_4$ the field dependence of $\lambda_d$ is slower but still appreciable: $\lambda_d(\mu_0H_L) \approx 0.1\lambda_d({\sim}\text{1~mT})$ at $\mu_0H_L \approx 100$~mT, where the increase of $\eta_d$ is just beginning. This demonstrates that for both compounds the suppression of $\lambda_d$ is not due to decoupling by $H_L$, which is significant only as $\eta_d \to 1$.\cite{HUIN79} It is also not due to the glassy dynamics mechanism of Keren \textit{et al.},\cite{KMCL96} which is not applicable when $\mu_0H_L \ll B_\mathrm{loc}$. The suppression occurs for fields two to three orders of magnitude smaller than $k_BT^*/\mu_B$, which is far too small to reduce the $3d$-spin fluctuation amplitude by simple alignment of the $3d$ spins. A shift of spectral noise power to higher frequencies is required, which must be due to some less direct mechanism. The nature of this mechanism, and why it results in a power-law field dependence for FeGa$_2$S$_4$, are unknown.

In NiGa$_2$S$_4$ $\eta_d$ increases from its low-field value for $\mu_0H_L \gtrsim \omega_\mu(0)/\gamma_\mu \approx 250$~mT, where $\omega_\mu(0)$ is the observed spontaneous muon precession frequency at $T = 0$.\cite{YDdRCM08,MNNH08} An increase is expected at precisely this field, as $H_L$ decouples the muon spin polarization from the quasistatic field.\cite{HUIN79} This quantitative agreement is strong evidence that the two-component structure of the muon relaxation function is not associated with separation of magnetically distinct phases. In that case $\eta_d$ would be the relative fraction of the slowly-relaxing phase; it and the initial 100\% muon spin polarization would both be field-independent. The stretched-exponential form of the late-time relaxation function indicates inhomogeneity of the muon relaxation rate, but there is no evidence for magnetic phase separation.

Below 100 mT $\eta_d$ varies somewhat more with field in FeGa$_2$S$_4$ than in NiGa$_2$S$_4$. This behavior is not understood, but might involve the significantly broader local field distribution in FeGa$_2$S$_4$.

\section{Discussion} \label{sec:disc}

\subsection{$\bm{T > T^*}$: comparison with the CSS result} \label{subsec:above}

We first consider muon spin relaxation for $T \ge T^*$.\cite{[{A similar analysis of $^7$Li nuclear spin relaxation in the 2D THAFM~$\mathrm{Li_7RuO_6}$ is given in }] IMYN09} The CSS calculation of the probe-spin dynamic relaxation rate~$1/T_1$ due to spin fluctuations in a 2D frustrated quantum antiferromagnet, discussed in more detail in the Appendix, yields
\begin{equation} 
1/T_1 = C_\mathrm{CSS}\left(\frac{N_0A_0}{\hbar}\right)^2 \frac{\hbar}{\rho_s} \left(\frac{T}{T_0}\right)^3 \exp(T_0/T)
\label{eq:CSS94T1}
\end{equation}
in the renormalized-classical (RC) regime~$T \ll 2\pi\rho_s$, with
\begin{equation} 
T_0 = 4\pi\rho_s \,.
\label{eq:CSS94T0}
\end{equation}
Here $N_0A_0$ is a renormalized hyperfine coupling constant, \cite{CSS94b} $\rho_s$ is the spin stiffness constant, and $C_\mathrm{CSS}$ is a numerical constant that can be estimated from Eqs.~(\ref{eq:T1CSS}) and (\ref{eq:xi}) in the Appendix. It is equal to 114.59 if one assumes Eq.~(\ref{eq:T1CSS}) to be an equality, but this is an order-of-magnitude estimate. The theory assumes nearest-neighbor AFM coupling, so that for NiGa$_2$S$_4$, where third-nearest-neighbor interactions dominate,\cite{YKHN08,SJBN10} the parameters obtained are effective values.\cite{Chub12pc} 

In general $\rho_\parallel$ and $\rho_\perp$ of $\rho_s$ for spin twists parallel and perpendicular, respectively, to the 2D plane are not equal.\cite{CSS94c} In the present experiments the samples were randomly-oriented powders except for the mosaic of NiGa$_2$S$_4$ single crystals, for which $\lambda_d(T)$ is not very different from the powder.\cite{MNNH08} We therefore take $\rho_s$ in Eq.~(\ref{eq:CSS94T0}) to be the orientation average~$(2\rho_\parallel + \rho_\perp)/3$. 

\begin{figure}[t]
\includegraphics[clip=,width=8cm]{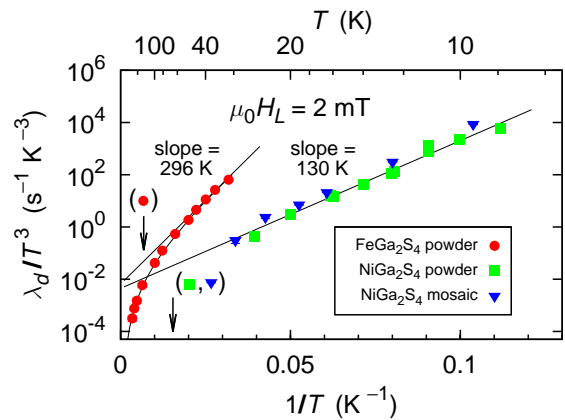}
\caption{\label{fig:CSS94} (Color online) Dependence of $\lambda_d/T^3$ on inverse temperature~$1/T$ in FeGa$_2$S$_4$ and NiGa$_2$S$_4$, $T \ge T^*$. Straight lines: fits of Eq.~(\protect\ref{eq:CSS94T1}). Labeled arrows: values of $1/2\pi\rho_s$ (see text). Curve: fit of the $Z_2$ vortex theory\protect\cite{KYO10,Kawa11} (Sec.~\protect\ref{sec:models}) to data for FeGa$_2$S$_4$.}
\end{figure}

We compare Eq.~(\ref{eq:CSS94T1}) to the muon relaxation rate~$\lambda_d$, recognizing that the uncertain absolute value of $1/T_1$ from Eq.~(\ref{eq:CSS94T1}) and the inhomogeneity make full agreement unlikely. Nevertheless the predicted $T^3\exp(T_0/T)$ temperature dependence might be expected. Figure~\ref{fig:CSS94} is a semi-log plot of $\lambda_d/T^3$ vs $1/T$ for both compounds at temperatures above $T^*$. 
For each compound the data extend down to the transition, i.e., the upper right-hand data points are for $T \approx T^*$. 

For FeGa$_2$S$_4$ the straight line is a fit of Eq.~(\ref{eq:CSS94T1}) to the lowest temperature points. For NiGa$_2$S$_4$ a straight line fits all the data for both powder and mosaic samples. The resulting values of $T_0$ yield $1/2\pi\rho_s = 2/T_0 = 0.0068~\mathrm{K}^{-1}$ and $0.0154~\mathrm{K}^{-1}$ for FeGa$_2$S$_4$ and NiGa$_2$S$_4$, respectively (arrows labeled by symbols in Fig.~\ref{fig:CSS94}), thereby justifying \textit{a posteriori} the procedure of obtaining $\rho_s$ from $T_0$.\cite{IMYN09} 

For NiGa$_2$S$_4$ all the data satisfy this condition, whereas for FeGa$_2$S$_4$ the data fall below the straight line at small $1/T$\@. This dropoff might be due to the onset of thermally-activated muon diffusion. It should be noted, however, that it occurs for $T \approx 2\pi\rho_s$, where the RC calculation ceases to be valid, and that similar behavior was observed in $^7$Li NMR in the THAFM~Li$_7$RuO$_6$ in the temperature range 20--40~K,\cite{IMYN09} where Li diffusion would not be expected. The situation is discussed further below. The curve is a fit of the $Z_2$ vortex theory\cite{KYO10,Kawa11} to the data, as described in Sec.~\ref{sec:models}.

The spin stiffness constants are expected to be proportional to the exchange constant~$J$; for a square lattice $\rho_s = JS^2$ in the classical limit.\cite{[{See, for example, }] SiHu89} For the 2D THAFM $\rho_\parallel$ and $\rho_\perp$ have been calculated in the spin-wave approximation, which yields\cite{CSS94c,LBLP95}
\begin{equation}
\frac{\rho_s}{JS^2} = \frac{(2/3)\rho_\parallel + (1/3)\rho_\perp}{JS^2} = \frac{1 - 0.399/2S}{\sqrt{3}}
\label{eq:rhoav}
\end{equation}
to first order in $1/S$\@. Equation~(\ref{eq:rhoav}) and $\rho_s$ obtained from the $\mu$SR data give a ``spin-wave'' value~$J_\mathrm{sw}$ of $J$, which can be compared with the value~$J_{\theta_W}$ obtained\cite{[{See, for example, }] Elli65} from $|\theta_{W}| = zJ_{\theta_W} S(S+1)/3$ for AF exchange between $z = 6$ near neighbors. 

Experimental values of these and other spin-system parameters and are given in Table~\ref{tab:CSS94} together with theoretical values of $\rho_s/J$.
\begin{table*}[t]
\caption{\label{tab:CSS94} Experimental and theoretical values of spin-system parameters in NiGa$_2$S$_4$ and FeGa$_2$S$_4$. $T_f$: freezing temperature from $\chi_\mathrm{ac}$ (NiGa$_2$S$_4$) and FC/ZFC bifurcation (FeGa$_2$S$_4$). $T^*$: transition temperature from $\mu$SR and ME data. $\theta_W$: (negative) Weiss temperature from $\chi(T)$. $\omega_\mu(0)$: spontaneous $T{=}0$ muon precession frequency (NiGa$_2$S$_4$ only). $N_0A_0/\hbar$, $T_0$: estimated renormalized muon hyperfine coupling constant (Ref.~\protect\onlinecite{CSS94b}) and characteristic temperature from fits of Eq.~(\protect\ref{eq:CSS94T1}) to $\lambda_d(T{\ge}T^*)$ (Fig.~\protect\ref{fig:CSS94}). $\rho_s$: orientation-averaged spin stiffness constant from Eq.~(\protect\ref{eq:CSS94T0}). $\rho_s/J$: ratio of $\rho_s$ to exchange constant~$J$ from calculation of Refs.~\protect\onlinecite{CSS94c,LBLP95}. $J_\mathrm{sw}$: exchange constant obtained from $\rho_s$. $J_{\theta_W}$: exchange constant obtained from $|\theta_W|$.}
\begin{ruledtabular}
\begin{tabular}{lccccccccccc}
& $S$ & $T_f$ & $T^*$ & $|\theta_W|$ & $\omega_\mu(0)$ & $N_0A_0/\hbar $ & $T_0$ & $\rho_s$ & $\rho_s/J$ & $J_\mathrm{sw}$ & $J_{\theta_W}$ \\
 & & (K) & (K) & (K) & ($10^6\ \mathrm{s}^{-1}$) & ($10^6\ \mathrm{s}^{-1}$) & (K) & (K) & & (K) & (K) \\ 
\colrule
NiGa$_2$S$_4$ & 1 & 2.2--2.7\footnotemark[1] & 9.0(0.5)\footnotemark[2] & 80(2)\footnotemark[3] & 200(10)\footnotemark[2] & 9.6(1.7) & 130(6) & 10.4(5) & 0.462 & 22(1) & 20.0(5) \\
FeGa$_2$S$_4$ & 2 & 16(1) & 31(2) & 160(9)\footnotemark[4] & $--$ & 63(9) & 296(18) & 23.5(1.4) & 2.079 & 11.3(7) & 13(1) \\
\end{tabular}
\end{ruledtabular}
\footnotetext[1]{Ref.~\onlinecite{Namb08}.}
\footnotetext[2]{Refs.~\onlinecite{YDdRCM08,MNNH08}.}
\footnotetext[3]{Ref.~\onlinecite{NNTS05}.}
\footnotetext[4]{Ref.~\onlinecite{NTON07}.}
\end{table*}
It can be seen that for both compounds the values of $J_\mathrm{sw}$ and $J_{\theta_W}$ are in good agreement. This is strong evidence that their spin dynamics above $T^*$ are dominated by the spin-wave fluctuations treated by CSS\@. For NiGa$_2$S$_4$ there is also reasonable agreement with the values~$J_3 = 32(7)$~K and 21~K of the dominant third-nearest-neighbor exchange constant obtained, respectively, from neutron scattering\cite{SJBN10} and the field dependence of the ESR resonance frequency well below $T^*$.\cite{YKHN08} The spin stiffness constants for the two compounds are in rough and good agreement with the values of $T^*$ for FeGa$_2$S$_4$ and NiGa$_2$S$_4$, respectively, as expected for Halperin-Saslow modes\cite{HaSa77} in spin-frozen states.\cite{PoKi09,NNO10} 

The renormalized muon-Ni$^{2+}$ hyperfine coupling constant~$N_0A_0/\hbar$ estimated from Eq.~(\ref{eq:CSS94T1}) is expected to be of the order of the spontaneous $T{=}0$ muon precession frequency~$\omega_\mu(0) = \gamma_\mu \langle B_\mathrm{loc}\rangle_\mathrm{av}(0)$, since the same interaction is involved in both cases. In FeGa$_2$S$_4$ $\omega_\mu(0)$ has not been measured (cf.\ Sec.~\ref{subsec:musr}), but in NiGa$_2$S$_4$ it is considerably larger than $N_0A_0/\hbar$ obtained from Eq.~(\ref{eq:CSS94T1}) (Table~\ref{tab:CSS94}). 

There are a number of uncertainties in the comparison, including the prefactor~$C_\mathrm{CSS}$ in Eq.~(\ref{eq:CSS94T1}) and the inhomogeneity in the relaxation. Perhaps the least certain quantity in the CSS expression for $1/T_1$ is the value of $\xi/c$,\cite{CSS94b} where $\xi$ is the spin correlation length and $c$ is the spin-wave velocity. Large discrepancies between calculated and observed correlation lengths have been reported.\cite{NNO10,KYO10} The situation is discussed in the Appendix. We conclude that the temperature dependence of $\lambda_d$ in NiGa$_2$S$_4$ is in good agreement with the CSS result, but the magnitude of the muon relaxation rate is not well understood. 

As shown in the Appendix, in the CSS treatment\cite{CSS94b} the spin-lattice relaxation rate falls below the RC temperature dependence in the crossover region between the RC ($T \ll 2\pi\rho_s$) and quantum-critical (QC) ($T \gg 2\pi\rho_s$) regimes. Thus the high-temperature dropoff in the relaxation rate in FeGa$_2$S$_4$ (Fig.~\ref{fig:CSS94}) might be due to this crossover. However, the QC regime, for which the condition~$\rho_s \ll J$ is necessary,\cite{CSS94b} is suppressed for large spin [cf.\ Eq.~(\ref{eq:rhoav})] and probably does not exist in FeGa$_2$S$_4$ ($S = 2$) because $\rho_s \approx 2J$ (Table~\ref{tab:CSS94}). Nevertheless, rapid non-universal short-range spin fluctuations may dominate at high temperatures\cite{CSS94b} and reduce $\tau_c$. The QC region might exist in NiGa$_2$S$_4$ ($S=1$), where $\rho_s \approx 0.5J$ (Table~\ref{tab:CSS94}), but here there is no sign of a dropoff for $T \lesssim 30$~K ($T/2\pi\rho_s \lesssim 0.46$); apparently the crossover has not been reached. Above $\sim$30~K $\lambda_d$ becomes too small to be measured reliably.

The ESR linewidth in NiGa$_2$S$_4$ above $T^*$ has been analyzed in terms of lifetime broadening by critical spin fluctuations.\cite{YKHN08,YKHN10} The linewidth depends much less on temperature than the muon relaxation rate, and exhibits an anomaly at $23~\mathrm{K} \sim 3\,T^*$ that was interpreted as a crossover between critical regimes. It should be noted that the ESR experiments were carried out in a field of 20~T, corresponding to a Zeeman energy~$\sim 2\,k_BT^*$; such a large field is expected to perturb the spin fluctuation spectrum in theoretical scenarios for the 2D THAFM.\cite{KYO10,SMSS11} It is also possible that the paramagnetic-state ESR line is statically broadened, since neutron scattering in NiGa$_2$S$_4$\cite{NNTS05,SJBN10} and muon relaxation in both compounds suggest inhomogeneous spin structure above $T^*$ that could be reflected in the susceptibility. Measurements of the ESR linewidth field dependence in the paramagnetic state might clarify the situation. There is no evidence for a crossover above $T^*$ in the muon relaxation data for NiGa$_2$S$_4$, which agree quantitatively with Eq.~(\ref{eq:CSS94T1}) up to $\sim$7~K above the ESR anomaly temperature. 

 \subsection{$\bm{T < T^*}$: quasistatic relaxation} \label{subsec:below}

Quasistatic muon spin relaxation (the onset of nonzero $\langle \mathbf{B}_\mathrm{loc}\rangle$) is observed in $\mu$SR experiments below $T^*$ in both NiGa$_2$S$_4$ and FeGa$_2$S$_4$. An important issue is whether or not the quasistatic field itself is truly static or slowly fluctuating. We discuss this in terms of two limiting scenarios: (1)~a static $\langle \mathbf{B}_\mathrm{loc}\rangle$ together with a fluctuating component~$\delta \mathbf{B}_\mathrm{loc}(t)$, and (2)~slow fluctuations of $\langle \mathbf{B}_\mathrm{loc}\rangle$ as a whole. We designate these scenarios ``conventional" and ``unconventional", respectively, for reasons that will become apparent, and consider the correlation time~$\tau_c$ associated with $\langle \mathbf{B}_\mathrm{loc}\rangle$. If quasistatic relaxation is observed $\tau_c$ must at least be longer than $1/\lambda_d$ (and possibly infinite), since otherwise motional narrowing would result in a single muon relaxation function rather than the two-term form of Eq.~(\ref{eq:2cmpnt}).\cite{HUIN79}

In the conventional scenario the dynamic muon relaxation is due to transitions induced by $\delta\mathbf{B}_\mathrm{loc}(t)$ between muon spin Zeeman levels in a truly static $\langle \mathbf{B}_\mathrm{loc}\rangle$ ($\tau_c = \infty$). This is the normal situation in an ordered magnetic phase, where $\delta \mathbf{B}_\mathrm{loc}(t)$ is due to thermally-excited magnons. These give rise to motionally-narrowed relaxation, because the frequencies associated with spin-flip scattering of magnons are much higher than $\gamma_\mu \delta B_\mathrm{loc}$.\cite{Jacc65inbib} In the unconventional scenario the muon relaxation is adiabatic,\cite{[{See Ref.~\protect\onlinecite{YaDdR11}, Sec. 6.3, also }] Abra61p34,HUIN79} and $\tau_c \approx 1/\lambda_d \sim 10^{-7}$ and $10^{-6}$~s for NiGa$_2$S$_4$ and FeGa$_2$S$_4$, respectively, below $T^*$ [Fig.~\ref{fig:rlx20G}(a)]. This would be extremely slow on the time scale of the $3d$ exchange interaction (${\sim}10^{-12}$~s), and would correspond to a highly unusual ``spin-gel'' phase as has been discussed in the literature.\cite{KaYa07,NNO10,KYO10,Kawa11}

Unfortunately the $\mu$SR data do not distinguish between these scenarios, and the observed value of $1/\lambda_d$ is therefore only a lower bound on $\tau_c$. Strong muon relaxation and NQR signal wipeout\cite{TIKI08} below $T^*$ could also occur in the conventional scenario.\footnote{The emergence of the NQR signal below $\sim$3~K  (Ref.~\onlinecite{TIKI08}) is not necessarily a sign of a phase transition, but simply means the spin-echo decay time~$T_2$ has become long enough so that the NQR signal is visible after the spectrometer ``dead'' time. This is not surprising, since $T_2 \propto 1/T$ at lower temperatures.} Nevertheless an extended critical regime or spin-gel phase, where constituent spins are strongly correlated but still fluctuating on the microsecond time scale with zero long-time average, is compatible with the $\mu$SR data for $T < T^*$. The absence of a sharp anomaly in bulk properties at $T^*$ would be highly unusual in the conventional scenario but is perhaps not ruled out in the unconventional scenario; this feature may be evidence in favor of the latter. More work is needed to clarify the situation.

As noted in Sec.~\ref{sec:intro}, the observation of the transition at $T^*$ in $^{57}$Fe ME experiments is evidence against the conjecture\cite{Varm08} that the muon electric charge perturbs its environment and somehow induces the transition observed in $\mu$SR\@. In NiGa$_2$S$_4$ substitutional Fe$^{2+}$ and host Ni$^{2+}$ ions are isovalent, and in $\mathrm{Ni_{0.99}Fe_{0.01}Ga_2S_4}$ the ME data yield a value of $T^*$ close to that from $\mu$SR in the undoped end compound.\cite{MKKK11} Fe$^{2+}$ is of course itself the host in FeGa$_2$S$_4$.

\subsection{Comparison with other models} \label{sec:models}

We compare the $\mu$SR results in FeGa$_2$S$_4$ and NiGa$_2$S$_4$ with candidate models for 2D THAFM behavior other than CSS.

\medskip \textit{$Z_2$ vortex binding-unbinding transition.} Kawamura and co-workers\cite{KaMi84,KaKi93,KaYa07,KYO10,Kawa11} and others\cite{HiKa86} have carried out extensive studies of $Z_2$ vortices as topological defects in a 2D THAFM with nearest-neighbor interactions. $Z_2$ vortices been studied in detail only in the classical limit, but presumably they also exist for quantum spins. It was suggested that binding of thermally-excited $Z_2$ vortices with decreasing temperature could lead to an unusual thermodynamic phase at a transition temperature~$T_v$, with long but finite spin correlation lengths and times below $T_v$.\cite{KYO10,Kawa11} 

It has been conjectured\cite{KaYa07,NNO10,KYO10,Kawa11} that the onset of quasistatic muon spin relaxation at $T^*$ might arise from this transition, so that $T^* = T_v$. The existence of a critical regime between $T_f$ and $T^*$ has been proposed for NiGa$_2$S$_4$ with $T_f/T^*\sim0.4$ (Refs.~\onlinecite{Namb08,NNO10}) and for NaCrO$_2$ with $T_f/T^*\sim0.3$.\cite{OMBU06} From the present results in FeGa$_2$S$_4$ $T_f/T^* \approx 0.5$, although we note again that the signature of $T_f$ is very different in the susceptibilities of FeGa$_2$S$_4$ and NiGa$_2$S$_4$. The observed rapid muon spin relaxation below $T^*$ and the lack of a divergence of $\lambda_d$ at $T^*$ seem consistent with an important qualitative feature of the $Z_2$ vortex picture, viz., finite spin correlation times at and below $T_v$. 

A number of questions remain concerning the $\mu$SR results and predictions of the $Z_2$ vortex binding model: 
\begin{itemize} \parindent1.5em

\item In the model $T_v$ is slightly lower than the temperature of the lower specific heat peak, as is the case for $T^*$ in NiGa$_2$S$_4$.\cite{DdRYCM09} As previously noted,\cite{DdRYMZ12} however, $T^*$ in FeGa$_2$S$_4$ lies between the two specific heat peaks, leading to doubt as to the applicability of the model to this compound. 

\item The theory predicts only weak essential singularities at $T_v$, where the fluctuation spectrum is dominated by the spin-wave contribution\cite{KYO10,Kawa11} so that the CSS calculation should apply. This varies exponentially (i.e., rapidly) with temperature, but a further abrupt increase of $\tau_c$ on cooling through $T^*$ (Ref.~\onlinecite{NNO10}) does not seem to be found in the $Z_2$ vortex picture; $\tau_c$ merely increases (exponentially) with cooling through $T_v$.\cite{KYO10,Kawa11} In this case the maximum rate is obtained for $\omega_\mu\tau_c \approx 1$ (i.e., not necessarily at $T_v$),\cite{HUIN79} and is of the order of $\omega_\mu \gtrsim 200\ \mu\mathrm{s}^{-1}$, whereas the observed maximum rates in FeGa$_2$S$_4$ and NiGa$_2$S$_4$ [Fig.~\ref{fig:rlx20G}(a)] are one to two orders of magnitude slower than this. Furthermore, such a slowing-down would not account for the mean-field-like temperature dependence of $\omega_\mu(T)$ below $T^*$ observed in NiGa$_2$S$_4$,\cite{YDdRCM08,MNNH08,MKKK11} since there would be no temperature dependence once $\omega_\mu\tau_c \gtrsim 1$. Thus the data seem to rule out a smooth increase of $\tau_c$ on cooling with no other effects.

\item In the $Z_2$ vortex model, spin dynamics above $T_v$ involve both unbound vortices and conventional spin-wave excitations. A crossover from spin-wave-dominated to vortex-dominated spin dynamics with increasing temperature is predicted at a temperature slightly above $T_v$.\cite{KYO10,Kawa11} No such crossover or other sign of spin dynamics beyond the CSS prediction is observed in the $\mu$SR data for NiGa$_2$S$_4$. As noted above a candidate for a crossover has been observed in ESR experiments on this compound,\cite{YKHN08,YKHN10} but the applied field used in these experiments (20~T) was much greater than the value ($\sim$0.1~T) sufficient to change the symmetry of the Hamiltonian and suppress $Z_2$ vortices.\cite{KYO10} 

\item An exponentially growing $Z_2$ vortex density at high temperatures leads to a decrease in $\tau_c$.\cite{KYO10,Kawa11} The resultant motional narrowing could explain the observed high-temperature dropoff of $\lambda_d/T^3$ in FeGa$_2$S$_4$ (Fig.~\ref{fig:CSS94}) below the CSS prediction [Eq.~(\ref{eq:CSS94T1})]. The curve in Fig.~\ref{fig:CSS94} is a fit to the FeGa$_2$S$_4$ data of an expression of the form
\begin{equation}
\lambda_d = A\frac{\tau_\mathrm{sw}\tau_v}{\tau_\mathrm{sw} + \tau_v}
\end{equation}
suggested by the $Z_2$ vortex model, where $\tau_\mathrm{sw}$ and $\tau_v$ are the spin correlation times associated with spin-wave and $Z_2$-vortex fluctuations, respectively.\cite{KYO10} Each correlation time incorporates its predicted temperature dependence, i.e., $\tau_\mathrm{sw} \propto T^3\exp(T_0/T)$ and $\tau_v \propto \exp\left\{[T_{Z_2}/(T-T_v)]^\alpha\right\}$, $\alpha \approx 0.5$.\cite{KYO10,Kawa11} The fit is good (curve in Fig.~\ref{fig:CSS94}), but the parameters are not well determined from the fit:  $T_{Z_2} = (1200 \pm 2600)$~K and $\alpha = 0.5\pm 1.8$. As noted in Sec.~\ref{subsec:above}, however, the dropoff in FeGa$_2$S$_4$ may be due to other mechanisms, and cannot be unambiguously attributed to muon relaxation by $Z_2$ vortices. 

\end{itemize}

Thus there are points of agreement and points of disagreement between the $Z_2$-vortex model and the experimental $\mu$SR results. Additional work is necessary to resolve these issues. In particular, a quantitative estimate of the expected muon relaxation rate due to unbound $Z_2$ vortices above $T_v$ is necessary to determine whether the negative result in NiGa$_2$S$_4$ would be expected. 
 
\medskip \textit{Spin-nematic ground state, impurity spin dynamics.} Treatments of the unconventional properties of NiGa$_2$S$_4$ via ``spin nematic'' models with nonmagnetic quadrupolar ground states\cite{LMP06,TsAr07,LZS07,LZN10,TaTs11} are in agreement with the field-independent specific heat, but appear to be called into question by the observation of strong spin dynamics at low temperatures.\cite{NNO10} These theories have specifically considered only $S = 1$ systems,\cite{[{A ``tensor magnet'' ground state has been proposed for $S = 2$: }] CIIK91} and therefore do not explicitly address the similar behavior of the two compounds. However, the observation that the field independence and low-temperature $T^2$ behavior of the specific heat is preserved for integer-spin impurities in NiGa$_2$S$_4$ but not for half-integer-spin impurities\cite{NNMO08,NaNa11} indicates the importance of integer spin for these properties. 

A recent theory by Takano and Tsunetsugu\cite{TaTs11} concludes that bond disorder generates magnetic impurities in an antiferroquadrupolar\cite{TsAr06} spin-nematic state of a $S{=}1$ 2D THAFM with biquadratic interactions. Coupling to nonmagnetic excitations gives rise to an indirect long-range interaction between these impurities. ``Freezing'' of impurity spins occurs to a state with slowly-fluctuating spin moments with vanishing spin expectation value. The time scale of these fluctuations has not been reported. It is argued that slow fluctuations are also induced in the lattice spins, leading to a fluctuating field throughout the system for impurity concentrations of the order of 1\%. Vortex unbinding strongly suppresses the impurity-spin interactions by preventing the definition of a consistent spin quadrupole director over a path around a vortex. This strong interplay between impurity spins and vortices leads to identification of $T_v$ as the transition temperature for impurity-spin-moment freezing as well as vortex binding.

Like the $Z_2$-vortex scenario, the spin-nematic impurity model is qualitatively consistent with a number of features of the $\mu$SR data. The abruptness of the transition, the possibility that the fluctuations are slow, and their suppression by relatively low magnetic fields coexisting with a field-independent $T^2$ specific heat are all consistent with the $\mu$SR results. Impurities give rise to magnetic inhomogeneity in the surrounding lattice, which might account for the observed sub-exponential muon relaxation. 

The nature of the impurity-spin fluctuations in this model is an important question. They appear to be quantum fluctuations, reflecting the vanishing of the expectation value of the impurity spin in the ground state. One must be cautious in assuming that such ground-state fluctuations will result in probe-spin relaxation at a neighboring site, however, since the ground state is by definition an eigenstate that does not fluctuate; thermal excitations are generally required for spin-probe relaxation in magnetic systems. 

As a simple example, consider two anti\-ferro\-magnet\-ic\-ally-coupled spins~$A$ and $B$, where a probe spin is hyper\-fine-coupled only to spin~$A$\@. One might expect the quantum fluctuations (oscillations) of $A$ to induce transitions between probe-spin states. It is straightforward to show, however, that the matrix elements of the spin operator~$\mathbf{S}_A$ in the hyperfine coupling vanish in the singlet ground state, so that the probe spin is a constant of the motion even though the spin of $A$ is not. It would be useful to clarify whether impurity-spin fluctuations in the spin-nematic impurity model are similarly ineffective at probe-spin relaxation or, alternatively, involve thermal excitation of degenerate or nearly-degenerate spin states. In the latter case persistent probe-spin relaxation at low temperatures would be expected.

There are other questions concerning comparison of this theory with the experimental situation:
\begin{itemize} \parindent1.5em

\item Consideration of a number of properties of NiGa$_2$S$_4$ within the theory has not yet been reported. These include a quantitative estimate of the impurity-spin fluctuation rate, its temperature dependence, the origin and effect of spin freezing at $T_f < T^*$, and the absence of persistent low-temperature relaxation in NQR experiments,\cite{TIKI08} which is in marked contrast with its presence in the $\mu$SR results.

\item Defect concentrations in NiGa$_2$S$_4$ have been found to be much lower than 1\%,\cite{NMHI09,NNO10} and $\mu$SR experiments have been carried out on a number of samples of NiGa$_2$S$_4$ from several laboratories with similar results.\cite{YDdRCM08,MNNH08,MNNI09,DdRYCM09,MNOM10} This suggests that impurities might not play a major role unless the concentration dependence is expected to be weak. On the other hand, the observation of two resonances in gallium NQR experiments on NiGa$_2$S$_4$\cite{TIKI08} suggests structural inhomogeneity at some level; stacking faults have been suggested. 

\item The identification of impurity-spin freezing with vortex binding below $T_v$ is based on the presence of unbound quadrupolar vortices above this temperature.\cite{TaTs11} As discussed above, there is no evidence for unbound $Z_2$ vortices above $T^*$ from $\mu$SR in NiGa$_2$S$_4$, and it has been argued\cite{KaYa07} that $Z_2$ vortices based on noncollinear AF order are the most likely species in this compound. Muon spin relaxation by nonmagnetic quadrupolar vortex excitations would, however, be weak and likely to be masked by relaxation due to spin fluctuations. 

\end{itemize}
We conclude that the spin-nematic model with impurity-spin freezing is also a promising candidate mechanism, but that more work is required to determine the extent to which it is supported by $\mu$SR and other experiments. One important aspect is the need for evidence for or against the substantial biquadratic spin interaction that is necessary for a nematic ground state.\cite{TaTs11}

\medskip \textit{Broken $C_3$ symmetry.} Models with nearest- and third-nearest-neighbor exchange couplings, with\cite{STB09} or without\cite{TaKa08} biquadratic coupling, exhibit ground states with broken $C_3$ rotational symmetry. Growth of quadrupolar short-range order at high temperatures, proposed\cite{STB09} to account for the high-temperature specific heat peak, is not easily tested by the $\mu$SR data, since as noted above quadrupole excitations would be hard to detect with a magnetic probe such as the muon spin. The magnetic correlations that generate $T^*$ are associated with the lower specific heat peak,\cite{STB09} however, whereas in FeGa$_2$S$_4$ $T^*$ is well above this peak.\cite{DdRYMZ12} 

\section{Conclusions} 

$\mu$SR studies of FeGa$_2$S$_4$ indicate drastic slowing and possible freezing of magnetic fluctuations below an unconventional transition at $T^* \approx 31$~K, which is twice the spin glass-like freezing temperature $T_f \approx 16$~K from magnetic susceptibility measurements. Muon spin relaxation rates above $T^*$ in both FeGa$_2$S$_4$ and NiGa$_2$S$_4$ are in very good quantitative agreement with the CSS result\cite{CSS94b} for spin-lattice relaxation in a 2D quantum antiferromagnet\@. An extended regime of strong spin fluctuations is observed for $T < T^*$, and the spin dynamics scale between the compounds from ${\sim}T_f$ to above $T^*$. 

Although there are differences in the magnetic properties of FeGa$_2$S$_4$ and NiGa$_2$S$_4$ (most notably the very different anomalies at $T_f$), similarities in the $\mu$SR data include the 2D critical spin dynamics above $T^*$, the absence of a divergence of $\lambda_d$ at $T^*$, and scaling between the two compounds in the neighborhood of $T^*$. These features are qualitatively consistent with theories of phase transitions driven by two distinct types of defects: (1)~$Z_2$-vortex topological excitations,\cite{KYO10,Kawa11} and (2)~magnetic impurities in a nonmagnetic spin-nematic ground state.\cite{TaTs11} There are questions concerning reconciliation of either model with other aspects of the data, however. The few differences in the $\mu$SR behavior of the two compounds include a possible anomaly in $\lambda_d(T)$ at $T_f$ in NiGa$_2$S$_4$ [Fig.~\ref{fig:rlx20G}(a), but see the associated discussion] and the dropoff of $\lambda_d(T)/T^3$ at high temperatures in FeGa$_2$S$_4$ (Fig.~\ref{fig:CSS94}). 

It should be noted that in general an exponential temperature dependence is expected from any large-scale excitation with constant energy.\cite{Kawa12pc} As discussed in Sec.~\ref{subsec:above}, the strongest evidence that in FeGa$_2$S$_4$ and NiGa$_2$S$_4$ muon relaxation is dominated by the specific spin-wave excitations treated by CSS is the remarkable agreement of the $\mu$SR values of the exchange constants with those from other experiments (Table~\ref{tab:CSS94}). 

The $\mu$SR experiments do not discriminate between slow fluctuations and truly static freezing below $T^*$. Nevertheless, a sharp anomaly is virtually always seen in some bulk property at a magnetic transition. In FeGa$_2$S$_4$ the absence of an anomaly at $T^*$ in the susceptibility and at $T_f$ in the muon spin relaxation appears to rule out a transition to a fully spin-frozen state at either of these temperatures. This and the coexistence of strong and strongly field-dependent magnetic spin dynamics with a field-independent specific heat at low temperatures are probably the two most remarkable results of this study.

\begin{acknowledgments}
We are grateful for assistance with the experiments from the staff of the TRIUMF Center for Molecular and Materials Science. One of us (D.E.M.) wishes to thank L. Balents, Hu Cao, A.~V. Chubukov, H. Kawamura, R.~R.~P. Singh, and C.~M. Varma for helpful correspondence and discussions. This work was supported by the U.S. NSF, Grants 0604105 and 1105380 (CSULA) and 0801407 (UCR), by the European Science Foundation through the Highly Frustrated Magnetism program, and by the Japan MEXT, Grants-in-Aid Nos.~17071003, 19052003, and 24740223. 
\end{acknowledgments}

\appendix* \section{Spin-lattice relaxation in the renormalized-classical and quantum-critical regimes of a 2D frustrated antiferromagnet}

We consider the CSS calculation of the spin-lattice relaxation rate~$1/T_1$ (Ref.~\onlinecite{CSS94b}) in more detail, and compare their results in the renormalized-classical (RC) and quantum-critical (QC) regimes of a 2D frustrated antiferromagnet that orders at $T = 0$. We follow the equation numbering and notation of CSS. 

In the RC region 
\begin{equation}
\frac{1}{T_1^\mathrm{(RC)}} \propto \left(\frac{A_0}{\hbar}\right)^2 \frac{N_0^2\xi}{c} {\left(\frac{k_BT}{\rho_s}\right)^{7/2}}\kern-10pt, \quad k_BT \ll 2\pi\rho_s\,, \tag{4.29}
\label{eq:T1CSS}
\end{equation}
where $A_0$ is the hyperfine coupling constant, $N_0$ is the condensate magnitude (renormalized by quantum fluctuations but presumably of order unity), $\xi$ is the correlation length, $c$ is the spin-wave velocity, and $\rho_s$ is the spin-wave stiffness constant. The prefactor required to make Eq.~(\ref{eq:T1CSS}) an equality is of order unity.\cite{Chub12pc} The correlation length in the RC region is given by
\begin{equation}
\xi = \frac{1}{2}\, \overline{\xi}\, \frac{\hbar c}{k_BT} \left(\frac{(N{-}1)k_BT}{4\pi\rho_s}\right)^{1/2(N{-}1)} \nonumber \exp\left[\frac{4\pi\rho_s}{(N{-}1)k_BT}\right] \,, \tag{4.7}
\label{eq:xi}
\end{equation}
where $N=2$ is the dimensionality and
\begin{equation}
\overline{\xi} = \left({\textstyle\frac{1}{8}}e\right)^{1/2(N{-}1)} \times \Gamma (1+1/2(N{-}1)) \tag{4.8}
\end{equation}

\vspace{-20pt} \begin{equation}
\hspace{56pt} = \left({\textstyle\frac{1}{8}}e\right)^{1/2} \times \Gamma (3/2) = \left({\textstyle\frac{1}{8}}e\right)^{1/2}\frac{\sqrt{\pi}}{2} = 0.5166 \,.
\end{equation}

The CSS result for $1/T_1^{\mathrm{(RC)}}$ is proportional to the ratio~$\xi/c$. In Sec.~\ref{subsec:above} we note that the discrepancy between the fit value of $N_0A_0/\hbar$ in NiGa$_2$S$_4$ and the expected value~${\sim}\omega_\mu(0)$ might be related to uncertainty in this quantity. Neutron scattering experiments in NiGa$_2$S$_4$\cite{SJBN10} obtained $\xi = 25$~\AA\ at 10~K and spin-wave velocity~$\hbar c = 29$~meV~\AA\ at 1.5~K, yielding $\xi/c = 5.7 \times 10^{-13}$~s. Using the value of $\rho_s$ for NiGa$_2$S$_4$ from Table~\ref{tab:CSS94}, the calculated CSS value of $\xi/c$ from Eq.~(\ref{eq:xi}) at $T^* = 9$~K is $1.04 \times 10^{-7}$~s, more than five orders of magnitude larger than the neutron scattering value. Assuming $N_0A_0/\hbar \approx \omega_\mu(0)$, the relaxation data and Eq.~(\ref{eq:T1CSS}) yield $\xi/c \approx 2.5  \times 10^{-10}$~s, more than two orders of magnitude smaller than that from Eq.~(\ref{eq:xi}) but still much larger than the neutron scattering value. Clearly this parameter is not well understood at present.

In the QC region the CSS result for $1/T_1$ is
\begin{equation}
\frac{1}{T_1^\mathrm{(QC)}} = \left(\frac{A_0}{\hbar}\right)^2 Z\, \frac{N_0^2\hbar}{\rho_s} \left(\frac{Nk_BT}{4\pi\rho_s}\right)^{\overline{\eta}}\kern-6pt , \quad k_BT \gg 2\pi\rho_s \,, \tag{5.15}
\end{equation}
where $Z = (\sqrt{5}-1)/8N = 0.07725$ for $N = 2$, and $\overline{\eta} = 1 + 32/3\pi^2N$ is a scaling exponent $= 1.5404$ for $N = 2$. 

To compare the results in the RC and QC regions we define the dimensionless quantities
\begin{equation}
\frac{1}{t_1} = \frac{\rho_s}{\hbar} \frac{1}{N_0^2}\left(\frac{\hbar}{A_0}\right)^2\frac{1}{T_1} \,, \quad t = k_BT/2\pi\rho_s \,,
\end{equation}
so that
\begin{eqnarray}
\frac{1}{t_1^{\mathrm{(RC)}}} & \propto & \frac{1}{2}\,\overline{\xi}\,(2\pi)^{5/2}\, 2^{-1/2}\,t^3 \exp(2/t) \nonumber \\
& = & 18.074\, t^3 \exp(2/t)\,, \quad t \ll 1 \,, \label{eq:RC}
\end{eqnarray}
and
\begin{eqnarray}
\frac{1}{t_1^{\mathrm{(QC)}}} & = & Zt^{\overline{\eta}} \nonumber \\
& = & 0.07725\, t^{1.5404} \,, \quad t \gg 1 \,. \label{eq:QC}
\end{eqnarray}

\begin{figure}[ht]
\includegraphics[clip=,width=8cm]{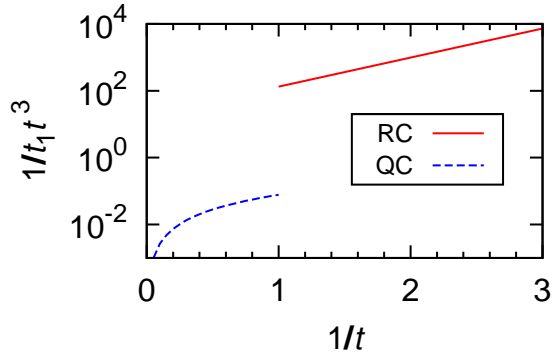}
\caption{\label{fig:A1} (Color online) dependence of $1/t_1t^3$ on $1/t$, where $1/t_1$ is the reduced dynamic relaxation rate from Ref.~\protect\onlinecite{CSS94b} [Eqs.~(\protect\ref{eq:RC}) and (\protect\ref{eq:QC})] and $t$ is the reduced temperature (see text for definitions). Solid line: renormalized-classical (RC) region. Dashed curve: quantum-critical (QC) region.}
\end{figure}

In Fig.~\ref{fig:A1} $1/t_1t^3$ is plotted versus $1/t$ for each region all the way to $1/t = 1$ [i.e., beyond the validity of Eqs.~(\ref{eq:RC}) and (\ref{eq:QC})]. 
It can be seen that at $1/t = 1$ the extrapolated value of $1/t_1t^3$ is much larger for the RC regime than for the QC regime. Thus in an exact solution $1/t_1$ would be expected to drop below the asymptotic RC result [Eq.~(\ref{eq:RC})] as the QC region is approached, although the caveat concerning the prefactor in Eq.~(\ref{eq:T1CSS}) should be noted. 



%

\end{document}